 \documentstyle[12pt]{article}
\input amssym.def
\input amssym.tex

\def\goth{\frak}
\def\double{\Bbb}

\def\cc{{\double C}}

\def\rr{{\double R}}

\def\qqq{{\double Q}}

\def\aa{{\cal A}}

\def\gg{{\goth g}}
\def\hh{{\cal H}}
\def\hhh{{{\double H}}}

\def\aa{{\cal A}}

\def\hh{{\cal H}}

\def\lll{{\cal L}}

\def\t{\,{\rm tr}\,}

\def\ddee{{\,\hbox{${\rm D}\!\!\!\!/\,$}}}
\def\dee{\,\hbox{\rm D}}
\def\de{\,\hbox{\rm d}}

\def\lb{\left[}
\def\rb{\right]}

\def\ot{\otimes}
\def\op{\oplus}

\def\bb{\begin{eqnarray}}
\def\ee{\end{eqnarray}}
\def\eee{\nonumber\end{eqnarray}}
\def\pp{\pmatrix}
\def\qq{\quad}

\begin{document}

\hsize 17truecm
\vsize 24truecm
\font\twelve=cmbx10 at 13pt
\font\eightrm=cmr8
\baselineskip 18pt

\begin{titlepage}

\centerline{\twelve CENTRE DE PHYSIQUE THEORIQUE}
\centerline{\twelve CNRS - Luminy, Case 907}
\centerline{\twelve 13288 Marseille Cedex}
\vskip 4truecm

\centerline{\twelve Riemannian and
Non-commutative Geometry in Physics}

\bigskip

\begin{center}
{\bf Bruno IOCHUM}
\footnote{ and Universit\'e de Provence,
\qq\qq iochum@cpt.univ-mrs.fr
\qq schucker@cpt.univ-mrs.fr} \\
\bf Daniel KASTLER
\footnote{ and Universit\'e de la M\'editerrann\'ee} \\
\bf Thomas SCH\"UCKER $^{1}$
\end{center}

\vskip 2truecm
\leftskip=1cm
\rightskip=1cm
\centerline{\bf Abstract}

\medskip

We feel that non-commutative geometry is to particle
physics what Riemannian geometry is to gravity. We
try to explain this feeling.

\vskip 1truecm
PACS-92: 11.15 Gauge field theories\\
\indent
MSC-91: 81E13 Yang-Mills and other gauge theories

\vskip 2truecm

\noindent oct. 1995
\vskip 1truecm
\noindent CPT-95/P.3260\\
\noindent hep-th/9511011

\vskip1truecm

 \end{titlepage}

We feel that non-commutative geometry is as
fundamental to physics as Minkowskian and
Riemannian geometry. Let us try to explain this by
comparing the standard model of particle physics and
general relativity. From a chronological point of view,
this comparison is difficult, because Riemannian
geometry existed well before general relativity.
However, the field theoretic approach allows to
introduce general relativity in close analogy to
classical electrodynamics without use of Riemannian
geometry. Therefore this approach is well suited for
our comparison.

So let us imagine a world ignoring Riemannian
geometry where physicists try to describe gravity.
They are inspired by Maxwell who takes a field $A$ of
spin 1, a second order differential operator $D_{Max}$
and writes down his field equation
\bb D_{Max}\,A=\frac{1}{c^2\epsilon_0}\,j,\eee
where $j$ is the source, charge density and currents,
and $\epsilon_0$ is the proportionality constant from
Coulomb's law. After many ingenious and expensive
experiments and theoretical trials and errors, the
physicists agree on the standard model of gravity. It
starts from a particular spin 2 field $g$, and a second
order differential operator $D_{Ein}$. The field
equation is
\bb D_{Ein}\,g=-\frac{8\pi G}{c^4}\,T,\eee
where the source $T$ is energy-momentum density
and currents, and $G$ is the proportionality constant
from Newton's universal law of gravity. Although in
perfect agreement with experiment, this standard
model has draw backs: who ordered spin 2? Maxwell's
differential operator $D_{Max}$ contains 8 summands,
the gravitational one $D_{Ein}$ results from brute
force and contains roughly 80 000 summands. Some of
these summands still are unaccessible to experiment.
At this stage, Riemannian geometry is discovered, the
spin 2 field is recognized as the metric and the
differential operator $D_{Ein}$ is recognized as the
curvature if the unknown summands are chosen
properly. Most physicists say: so what, just fancy
mathematics. Some dream of a geometric unification
of all forces. Later, even more expensive experiments
will test the predictions of Riemannian geometry
coming from the unknown summands.

If, in the real world, we qualify general relativity as
revolution, we have several criteria.
\begin{itemize}
\item
Postdiction: the theory correctly reproduces
experimental data, that remain unexplained in the old
theories, e.g. the precession of perihelia of Mercury.
\item
Prediction: the theory can be in contradiction with
future experimental data, e.g.
deflection of light.
\item
New concepts, e.g. curved spacetimes, absence of
universal time.
\item
Reticence of the majority.
\end{itemize}

Our purpose is to explain that for non-commutative
geometry the analogue of $g$ in the imaginary world
is the Higgs field, the analogue of $D_{Ein}$ is the
Lagrangian of the standard model of electro-weak and
strong interactions. Postdictions of the theory are that
fermions sit in fundamental representations, that
weak interactions violate parity, that strong
interactions are vector like and
\bb&&\rho:=\frac{g_1^2+g_2^2}{g_2^2}\,
\frac{m^2_W}{m^2_Z}=1,\cr
\cr&& m_e<m_W<m_t/\sqrt3, \cr&&
2g_1^{-2}>g_2^{-2}+g_3^{-2}/3.\eee
There is also a prediction, the mass of the Higgs,
accessible to experiment in about ten years.
New concepts are fuzzy spacetimes --- that is
spacetimes with an uncertainty relation --- and
discrete spacetimes.

\section{The Establishment}

Let us briefly summarize today's established theory of
particles and interactions. It is a particular
Yang-Mills-Higgs theory. To get started, we view this
class of theories as black box or slot machine. The
input comes in two parts, bills and coins. The output is
a particle phenomenology, that is cross
sections, branching ratios, life times ... To decide
whether a particular input is a winner, its
corresponding output is confronted with millions
of experimental numbers that cost billions of Swiss
Francs.

\subsection{Bills and coins}

The Yang-Mills-Higgs machine has four slots
for one bill each. In the first of these slots you are
supposed to put a finite dimensional, real, compact Lie
group $G$. For the remaining slots choose three
unitary representations $\rho_L$, $\rho_R$,
$\rho_S$ defined on Hilbert spaces $\hh_L$, $\hh_R$,
$\hh_S$. These Hilbert spaces will accommodate the
left- and right-handed fermions and the Higgs scalars.

After having eaten these four bills, the
machine will ask you for coins, real or complex
numbers. The number of coins depends on the chosen
bills.
\begin{itemize}
\item
An invariant scalar product on the Lie algebra $\gg$
of $G$. This choice is parameterized by one positive
number $g$, the `gauge coupling', for every simple
factor in $G$, e.g.
\bb (b,b')&:=&\frac{1}{g_1^2}\bar bb',
\qq b,b'\in\ u(1),
\cr\cr  (a,a')&:=&\frac{2}{g_n^2}\t(a^*a'),\qq
a,a'\in su(n). \eee
\item
An invariant, positive polynomial $V(\varphi)$,
$\varphi\in\hh_S$ of order 4, the `Higgs potential'. We
want this potential to break $G$ spontaneously. This
means that no invariant vector in $\hh_S$ minimizes
$V$. For example if $G=SU(2)$ with the fundamental
representation $\hh_S=\cc^2$, the most general Higgs
potential is
\bb V(\varphi) = \lambda (\varphi^\ast
\varphi)^2 - {{\mu ^2 }\over 2}\varphi^\ast \varphi,
&&\qquad \varphi \in
\hh_S, \quad \lambda, \mu > 0.\eee
\item
One complex number or `Yukawa coupling' $g_Y$ for
every trilinear invariant --- i.e. for every one
dimensional invariant subspace, `singlet' --- in the
decomposition of the representation  associated to $
\left(\hh_L^{\ast}\otimes \hh_R\otimes \hh_S\right)
\oplus\left(
     \hh_L^\ast\otimes \hh_R\otimes \hh_S^*\right).
$
For example if $G=SU(2)$, $\hh_L=\cc^2$,
$\hh_R=\cc$, $\hh_S=\cc^2$ there is one singlet:
\bb \sum _{j=1}^2
\bar\psi_{Lj}\psi_R\varphi_{j},\qq
\pp{\psi_{L1}\cr \psi_{L2}}\in\hh_L,\
\psi_R\in\hh_R,\ \pp{\varphi_1\cr
\varphi_2}\in\hh_S.\eee
 \end{itemize}

Physicist have been playing on this slot machine for
the last thirty years. One winner clearly emerged,
the so called standard model. Its bills are
\bb G &= & SU(3) \times SU(2) \times U(1) \cr \cr
\hh_L &=& \bigoplus_1^3\lb (1,2,-\frac{1}{2})\oplus
(3,2,{1 \over 6}) \rb  \label{hl},\\
 \hh_R& = &\bigoplus_1^3\lb (1,1,-1)\oplus
(3,1,{2\over 3})\oplus (3,1,-{1\over 3}) \rb, \label{hr}
\\ \cr
 \hh_S &= &(1,2,-\frac{1}{2}) \label{hs},\ee
where $(n_3,n_2,y)$
denotes the tensor product of an $n_3$ dimensional
representation of $SU(3)$, an $n_2$ dimensional
representation of $SU(2)$ and the one dimensional
representation of $U(1)$ with hypercharge $y$:
\bb
\rho(e^{i\theta}) = e^{iy\theta}, &&\qquad
y\in\qqq,\   \theta \in [0,2\pi).\eee
Some vocabulary: particles are basis elements. The
spin 1 particles, the gauge bosons, span the Lie
algebra $\gg$ of the group $G$. The eight basis
elements of $su(3)$ are called gluons. They are
massless and mediate the strong interactions, e.g.
nuclear fusion, fission, $\alpha$-decay. The
remaining $su(2)\op u(1)$ is spanned by the photon
--- Maxwell's old friend and later found responsible for
$\gamma$-decay --- and three massive bosons, the
$W^+$, $W^-$ and $Z$. They mediate the weak
interactions, e.g. $\beta$-decay. The spin
$\frac{1}{2}$ particles or fermions come in three
identical copies, `generations'. The first generation of
$\hh_L$ is spanned  by the left-handed parts
(Weyl spinors) of the electronic neutrino, the electron
and the up and down quarks. The first two are called
leptons, from the greek word for mild, because sitting
in $SU(3)$ singlets they are not subject to strong
interactions. The other two left-handed generations
are spanned by  the muonic neutrino, the muon, the
charm and strange quarks, and the tau neutrino, the
tau, the top and bottom quarks. $\hh$ is spanned by the
right-handed parts of the same particles, except that
there are no right-handed neutrinos. Consequently
the neutrinos are massless.
 The particle
count for the spin 0 particles, scalars, is a little bit
more complicated. Not all basis elements of $\hh_S$
correspond to physical scalars. There is only one in
the standard model. It is called Higgs scalar and is still
being searched for.

 Because of the high
degree of reducibility in the bills, there are many
coins, among them 27 Yukawa couplings. Not all of
them have a physical meaning. They can be converted
into 18 physically significant, positive numbers
\cite{data}, three gauge couplings,
\bb g_1=0.3575\pm
0.0001,&g_2=0.6507\pm 0.0007,& g_3=1.207\pm
0.026,\eee eleven particle masses,  \bb m_W=80.22\pm
0.26\ {\rm GeV},&m_H>58.4\ {\rm GeV},\cr
m_e=0.51099906\pm 0.00000015\ {\rm MeV},&
m_u=5\pm 3\ {\rm MeV},&m_d=10\pm 5\ {\rm MeV},
\cr  m_\mu=0.105658389\pm 0.000000034\ {\rm GeV},&
m_c=1.3\pm 0.3\ {\rm GeV},&m_s=0.2\pm 0.1\
{\rm GeV},\cr
m_\tau=1.7771 \pm 0.0005\ {\rm GeV},&
m_t=176\pm 18\ {\rm GeV},&m_b=4.3\pm 0.2\
{\rm GeV},\eee
and quark mixings. These mixings are given in form
of a
unitary matrix, the Cabbibo-Kobayashi-Maskawa
matrix
\bb C_{KM}:=\pp{V_{ud}&V_{us}&V_{ub}\cr
V_{cd}&V_{cs}&V_{cb}\cr  V_{td}&V_{ts}&V_{tb}}.\eee
For physical purposes it can be parameterized by
three angles $\theta_{12}$,
$\theta_{23}$, $\theta_{13}$ and
one $CP$ violating phase $\delta$:
\bb C_{KM}=\pp{
c_{12}c_{13}&s_{12}c_{13}&s_{13}e^{-i\delta}\cr
-s_{12}c_{23}-c_{12}s_{23}s_{13}e^{i\delta}&
c_{12}c_{23}-s_{12}s_{23}s_{13}e^{i\delta}&
s_{23}c_{13}\cr
s_{12}s_{23}-c_{12}c_{23}s_{13}e^{i\delta}&
-c_{12}s_{23}-s_{12}c_{23}s_{13}e^{i\delta}&
c_{23}c_{13}},\eee
with $c_{kl}:=\cos \theta_{kl}$,
$s_{kl}:=\sin \theta_{kl}$.
The
absolute values of the matrix elements are:
\bb \pp{
0.9753\pm 0.0006&0.221\pm 0.003&0.004\pm
0.002\cr
0.221\pm 0.003&0.9745\pm 0.0007&0.040\pm 0.008\cr
0.010\pm 0.006&0.039\pm 0.009&0.9991\pm 0.0004}.
\eee
Every body agrees that the standard model is ugly, too
ugly to be a fundamental theory.

\subsection{The general rules}

Let us now have a closer look at the inside of the
Yang-Mills-Higgs machine. It produces a Lagrangian
that consists of five separate pieces. The first is the
{\it Yang-Mills Lagrangian}, well motivated
physically as non-abelian generalization of the
famous Maxwell Lagrangian, $G=U(1)$. Also on the
mathematical side, this Lagrangian needs no further
introduction. Its fundamental field are the gauge
bosons or connection, $A\in \Omega^1(M,\gg)$, a
1-form on the spacetime manifold $M$ with values in
the Lie algebra $\gg$:
\bb \lll_{YM}[A]=\frac{1}{4}(F,*F),\eee
where $F:=\de A+\frac{1}{2}[A,A]$ denotes the field
strength or curvature of $A$, $*$ is the Hodge star, and
$(\cdot,\cdot)$ is the chosen invariant scalar product
on $\gg$.

The second piece is the {\it Dirac
Lagrangian}. It is geometricly as noble as the
Yang-Mills Lagrangian.
  \bb
\lll_D[A,\psi_L,\psi_R]=
\psi_L^*\ddee\psi_L+\psi_R^*\ddee\psi_R,\eee
where $\psi_L$ is a left-handed spinor with values in
$\hh_L$, $\psi_L^*$ is its dual with respect to the
scalar product in $\hh_L$, $\ddee$ is the covariant
Dirac operator, $\dee \psi_L:=\de \psi_L+
\tilde\rho_L(A)\psi_L.$ We denote by $\tilde\rho$ the
Lie algebra representation belonging to the group
representation $\rho$. For $G=U(1)$ these two
Lagrangians yield the very successful quantum
electrodynamics and for $G=SU(3)$,
$\hh_L=\hh_R=\cc^3$ we get the present day theory
for strong interactions, quantum chromodynamics.

In order to incorporate weak interactions and to give
masses to gauge bosons and fermions, one is forced to
break the symmetry $G$ spontaneously. This is
where the patchwork starts. One has to
postulate the existence of
scalars, til now unobserved. They are 0-forms with
values in $\hh_S$,
\bb \varphi\in\Omega^0(M,\hh_S).\eee
 One also has to add three more Lagrangian pieces
involving the scalars, the {\it  Klein-Gordon
Lagrangian}
\bb \lll_{KG}[A,\varphi]=\frac{1}{2}
\dee\varphi^**\dee\varphi,\eee
the {\it Higgs potential} and the {\it Yukawa terms}
\bb \lll_{Yu}[\psi_L,\psi_R,\varphi]=
\sum_{j=1}^ng_{Yj}\left(\psi_L^*,\psi_R,\varphi
\right)_j+\sum_{j=n+1}^mg_{Yj}\left(\psi_L^*,\psi_R,
\varphi^*\right)_j\ +\ {\rm complex\ conjugate}.\eee

To summarize, the standard model has two major
shortcomings, the general rules of Yang-Mills-Higgs
look artificial, as well as the input, bills {\it and} coins,
singled out by nature. Nevertheless the standard model
has resisted to an extremely detailed experimental
analysis where all concurrent models have failed.

\section{The Revolution}

The non-commutative formulation improves the
situation on all three levels, general rules, bills and
coins.

\subsection{General rules}

The Yang-Mills and Dirac Lagrangians have a
geometric origin and Alain Connes found a natural
generalization of some of them to non-commutative
geometry \cite{cbook}. Connes and Lott have
considered these Lagrangians in the particular case of
a product geometry of an ordinary four dimensional
spacetime geometry by a zero dimensional
non-commutative geometry. There a miracle happens
\cite{cbook,cl}. When decomposing the
non-commutative versions of the Yang-Mills and Dirac
Lagrangians in terms of ordinary fields they retrieve
of course the ordinary Yang-Mills and Dirac
Lagrangians. Simultaneously and free of charge, they
also get the other three  pieces, the Klein-Gordon
Lagrangian, the symmetry breaking Higgs potential,
and some Yukawa terms. Every such Connes-Lott
model yields a particular Yang-Mills-Higgs model. The
contrary is far from being true, how far will be
discussed in terms of bills and coins in the following
subsections.

\subsection{Bills}

Since the introduction of quantum mechanics, we are
used to the description of non-commutative spaces in
terms of involution algebras. A zero dimensional
non-commutative space is given by a finite
dimensional, real involution algebra $\aa$. The group
$G$ of the ensuing gauge model will be the group of
unitaries of $\aa$
\bb \left\{a\in\aa|\ a^*a=a^*a=1\right\}\eee
or possibly a subgroup thereof. In order to construct
a differential calculus on the non-commutative space,
Connes introduces two algebra representations
$\rho_L$ and $\rho_R$ on Hilbert spaces $\hh_L$ and
$\hh_R$ such that $\rho_L\op\rho_R$ is faithful.
In the finite dimensional case, this implies that
$\aa=M_n(\rr)$, $M_n(\cc)$ or $M_n(\hhh)$, $\hhh$
denoting the quaternions, and that the algebra
representations are copies of the defining
representation. For $M_n(\cc)$, there is--- in addition
to the defining representation --- its conjugate. In
terms of bills of the resulting Yang-Mills-Higgs model
we have the following irreducible possibilities:
\bb G=O(n,\rr),&& \hh_{L,R}=\cc^n,\cr
G=U(n)\ {\rm or}\ SU(n),&&\hh_{L,R}=\cc^n,\cr
G=USp(n),&&\hh_{L,R}=\cc^{2n}.\eee
The restriction on the group bill is mild, only the
exceptional groups are excluded.  The restrictions on
the two fermionic bills is appreciable, e.g. $U(1)$ only
admits hypercharge -1, or 1, $SU(2)$ only has one
irreducible representation, the two dimensional one,
while in the general setting there is an infinite
number to choose from.

The restriction on the scalar
bill is spectacular. It comes out to be a {\it group}
representation, a unitary representation of the group
of unitaries, and is restricted by the
fermionic bills: its Hilbert space is an invariant
subspace,
\bb\hh_S\:\subset\:\left(\hh_L^*\ot\hh_R\right)\,\op\,
\left(\hh_R^*\ot\hh_L\right).\label{hsconstr}\ee
This invariant subspace is entirely determined by
the coins.

One is of course tempted to build models with a simple
algebra and/or irreducible fermion representations.
Besides phenomenological shortcomings, all such
models have a degenerate vacuum, an invariant
vector in $\hh$, that minimizes the Higgs potential.
All popular Grand Unifies Theories are excluded in
Connes and Lott's approach. Similarly, all left-right
symmetric models are excluded, because the
constraint (\ref{hsconstr}) forbids spontaneous parity
violation. The minimal non-commutative model
without degeneracy turns out to be the $SU(2)\times
U(1)$ model of weak interactions with {\it two}
generations of leptons:
\bb\hh_L &=& \bigoplus_1^2 (2,0)   ,\cr
 \hh_R& = &\bigoplus_1^2 (1,-1).\eee
Comparing with (\ref{hl}-\ref{hr}), we see that the
hypercharges are wrong. They are corrected by the
inclusion of strong interactions.

This inclusion requires a new ingredient \cite{creal},
a real structure or --- in physical terms --- a
generalization of charge conjugation to
non-commutative geometry. The existence of a real
structure implies additional constraints on the
fermion representations. The
representations (\ref{hl}-\ref{hs}) of the standard
model have four features.
\begin{itemize}\item
 The weak
interaction $SU(2)$ violates parity maximally, it acts
only on left-handed fermions.
\item
 The strong interaction
$SU(3)$ is vectorial, it acts in the same way on
left- and right-handed fermions.
\item
The scalars transform under $SU(2)$, implying
spontaneous breaking of $SU(2)$ that renders its
gauge bosons, $W^+$,  $W^-$ and $Z$, massive.
\item
The scalars do not transform under $SU(3)$. It
remains unbroken and its gauge bosons, the gluons,
massless.
\end{itemize}
In a Yang-Mills-Higgs theory these four features are
independent, not so in the non-commutative
approach. We already stated that the scalar
representation is not chosen and the two last features
follow from the two first. On top, the existence of a real
structure implies that the first feature implies the
second \cite{reb}.

The existence of a real structure is intimately related
to another mathematical property, a non-commutative
version of Poincar\'e duality which puts still another
constraint on the fermion representations. It turns
out that this constraint is fulfilled in the standard
model (\ref{hl}-\ref{hr}). However, slightly
modifying $\hh_R$ by adding right-handed neutrinos
--- a modification compatible with all constraints so
far \cite{gb} --- violates this additional constraint
\cite{creal}\cite{test}.

\subsection{Coins}

In a Yang-Mills-Higgs model, that comes from a
Connes-Lott model, the coins cannot be chosen
independently. In an arbitrary Yang-Mills-Higgs
model the choice of coins is a point in the space of
direct products of intervals. In a Connes-Lott model
this point must lie in a subspace. This subspace is a
submanifold with interesting structure. Depending on
the choice of bills, this submanifold may be of the
same dimension as its surrounding
hypercube or not. Due to the high degree of
reducibility of its fermionic Hilbert space, the
standard model is in the first case. Its Connes-Lott
submanifold is an open subset of its Yang-Mills-Higgs
hypercube given by the inequalities
\bb&& m_e<m_W<m_t/\sqrt3, \cr&&
2g_1^{-2}>g_2^{-2}+g_3^{-2}/3\cr &&
m_{H\,min}<m_H<m_{H\,max}.\eee
The bounds on the Higgs mass are complicated
functions of the other coins. The fact that the
non-commutative constraints on the parameters of
the standard model are inequalities rather than
equations may be important to insure their
stability under renormalization flow. On the other
hand, for the experimental values of the parameters
\bb\frac{m_{H\,max}-m_{H\,min}}
{({m_{H\,max}+m_{H\,min})/2}}\simeq
\frac{m_\tau^2-m_e^2}{m_t^2}\simeq 10^{-4}\eee
and for all practical purpose the Higgs mass is fixed.
To our knowledge, this is the first mass relation, that
comes with a (small) conceptual uncertainty and we
call it a fuzzy relation. We stress that the fuzziness of
the Higgs mass requires the existence of at least two
generations.

\section{Conclusion}

The first miracle of non-commutative geometry
applied to particle physics concerns the general
rules. Here, this geometry answers the question: Who
ordered the Higgs.

The two subspaces of bills and coins
accessible to a
Connes-Lott model have interesting structure and are
tiny compared to the two Yang-Mills-Higgs
hypercubes, fig. 1 and 2. To us, it is a second miracle
that the two points defining the standard model fall
into these tiny subspaces, at least as long as the Higgs
mass is unknown.

{\bf Figure captions}

\begin{description}
\item{Fig.1:}
 An artist's partial view of the space of bills of all
Yang-Mills-Higgs models and some of its subspaces.
$GUT$ stands for `Grand Unified Theories', i.e. $G$
simple. $L-R$ stands for left-right symmetric models,
i.e. $\hh_L=\hh_R$. $SM$ stands for standard model
and $CL$ for Connes-Lott models.
\item{Fig.2:}
Partial view of the space of coins of the
standard model,  lower and upper bounds of the
Higgs mass as a function of the top and $\tau$ masses,
all other coins are set to
their experimental values. For the experimental value,
$m_\tau\ =\ 1.8\ {\rm GeV}$, the two bounds differ by
 $10^{-2}\ {\rm GeV}$ in the indicated range of
$m_t$.

\end{description}

 
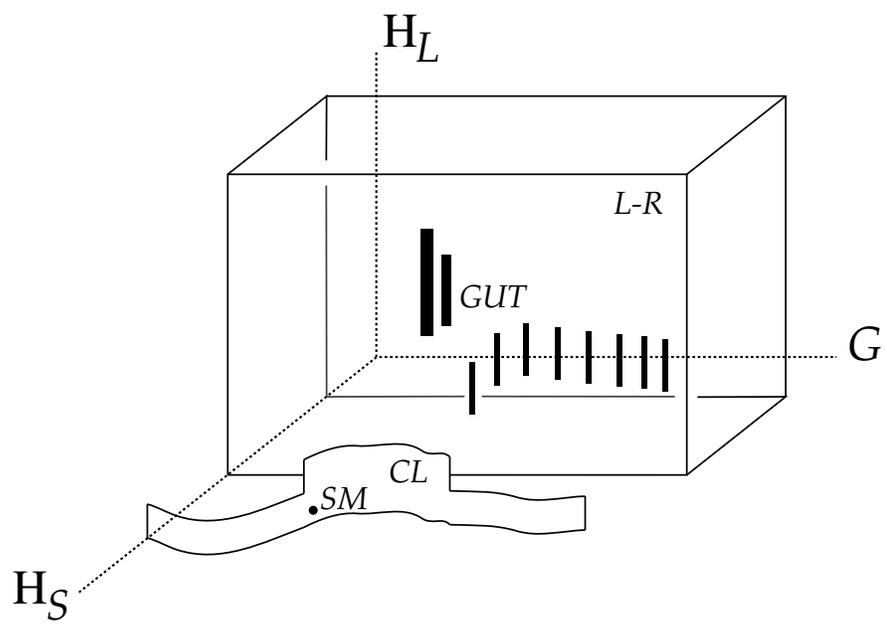
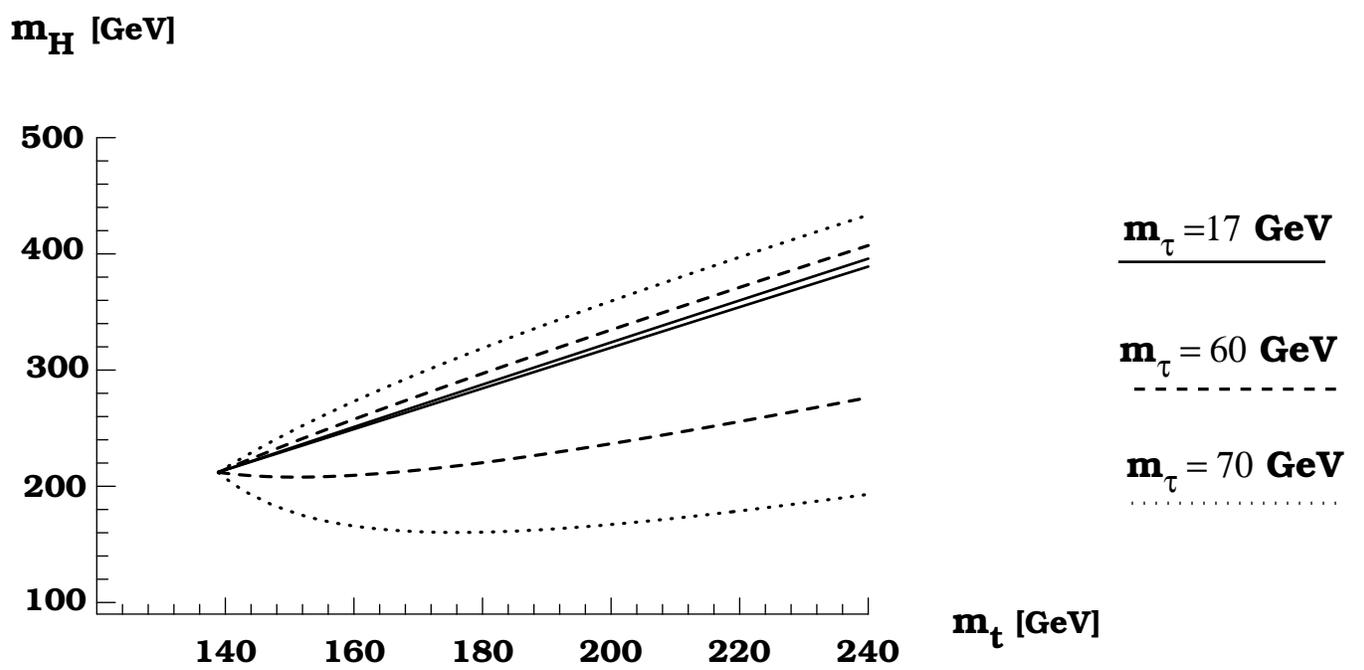
\end{document}